\documentclass[pre,preprint,showpacs,preprintnumbers]{revtex4}

\usepackage{graphicx}
\usepackage{latexsym}
\usepackage{amsmath}
\usepackage{amssymb}
\usepackage{amsfonts}
\usepackage{bm}

\newcommand{\la}{\left<}
\newcommand{\ra}{\right>}

\renewcommand{\vec}[1]{\ensuremath{\mathbf{#1}}}

\newcommand{\Rend}{\ensuremath{R_\text{e}}}
\newcommand{\Rgyr}{\ensuremath{R_\text{g}}}
\newcommand{\be}{\ensuremath{b_\text{e}}}
\newcommand{\bestar}{\ensuremath{b_{\text{e}\star}}}
\newcommand{\bg}{\ensuremath{b_\text{g}}}

\newcommand{\dc}{\ensuremath{d_\mathrm{p}}}

\newcommand{\Ge}{\ensuremath{G_\text{e}}}
\newcommand{\fRsN}{\ensuremath{\tilde{f}}}
\newcommand{\sprim}{\ensuremath{s^{\prime}}}
\newcommand{\Mmat}{\ensuremath{\underline{\underline{M}}}}
\newcommand{\trace}{\ensuremath{\text{tr}}}
\newcommand{\deter}{\ensuremath{\text{det}}}
\newcommand{\ginter}{\ensuremath{g_\text{inter}}}

\newcommand{\nchain}{\ensuremath{n_\text{chain}}}
\newcommand{\gcm}{\ensuremath{g_\text{cm}}}

\begin{document}

\title{Static Properties of Polymer Melts in Two Dimensions}

\author{H. Meyer}
\author{J.P. Wittmer}
\email{joachim.wittmer@ics-cnrs.unistra.fr}
\author{T. Kreer}
\author{A. Johner}
\author{J. Baschnagel}
\affiliation{Institut Charles Sadron, 23 rue du Loess, BP 84047, 67034 Strasbourg Cedex 2, France}

\date{\today}

\begin{abstract} 
Self-avoiding polymers in strictly two-dimensional ($d=2$) melts are investigated by means of
molecular dynamics simulation of a standard bead-spring model with chain lengths ranging up to $N=2048$. 
The chains adopt compact configurations of typical size $R(N) \sim N^{\nu}$ with $\nu=1/d$. 
The precise measurement of various distributions of internal chain distances allows a direct test of 
the contact exponents $\Theta_0=3/8$, $\Theta_1=1/2$ and $\Theta_2=3/4$ predicted by Duplantier.
Due to the segregation of the chains the ratio of end-to-end distance $\Rend(N)$
and gyration radius $\Rgyr(N)$ becomes $\Rend^2(N)/\Rgyr^2(N) \approx 5.3 < 6$ for $N \gg 100$
and the chains are more spherical than Gaussian phantom chains.
The second Legendre polynomial $P_2(s)$ of the bond vectors decays as 
$P_2(s) \sim 1/s^{1+\nu\Theta_2}$ measuring thus the return probability of the
chain after $s$ steps.
The irregular chain contours are shown to be characterized by a perimeter length 
$L(N) \sim R(N)^{\dc}$ of fractal line dimension $\dc = d-\Theta_2 =5/4$.
In agreement with the generalized Porod scattering of compact objects with fractal 
contour the Kratky representation of the intramolecular structure factor $F(q)$ 
reveals a strong non-monotonous behavior with $q^dF(q) \sim 1/(q R(N))^{\Theta_2}$ 
in the intermediate regime of the wave vector $q$. This may allow to confirm the 
predicted contour fractality in a real experiment. 
\end{abstract}

%
%

\maketitle

\section{Introduction}
\label{sec_intro}

%
Dense self-avoiding polymers in two dimensions (2D) have been considered 
theoretically \cite{DegennesBook,dupl,ANS03,jakob}, 
by means of computer simulation 
\cite{baumgartner82,carmesin90,ostrovsky97,rutledge97,pakula02,balabaev02,yethiraj03,yethiraj05,cavallo03,cavallo05,CMWJB05,MKA09}
and more recently even in real experiments \cite{mai99,wang04,deutsch05,dobrynin07,granick07,SB09}.
It is now generally accepted 
\cite{dupl,ANS03,jakob,baumgartner82,carmesin90,rutledge97,pakula02,balabaev02,yethiraj03,cavallo05,mai99,dobrynin07,MKA09}
(with the notable exception of Refs.~\cite{ostrovsky97} and \cite{wang04}) 
that these chains adopt compact and segregated conformations at high densities, 
i.e. as first suggested by de Gennes \cite{DegennesBook}
the typical chain size $R(N)$ scales as 
\begin{equation}
N \approx \rho R^d(N)
\label{eq_compactN}
\end{equation} 
with $N$ being the chain length, 
$\rho$ the monomer number density and $d=2$ the spatial dimension. 
We assume here that monomer overlap and chain intersections are strictly forbidden \cite{ANS03}.
Compactness and segregation are expected to apply not only on the scale 
of the total chain of $N$ monomers but 
also to subchains comprising $s$ monomers \cite{ANS03,dobrynin07}, 
at least as long as $s$ is not too small.
The typical size $R(s)$ of subchains should thus scale as 
\begin{equation}
R(s) \approx  (s/\rho)^{\nu} \mbox{ for } 1 \ll s \le N
\label{eq_compact}
\end{equation}
with a Flory exponent $\nu=1/d = 1/2$ set by the spatial dimension.  
Interestingly, the direct visualization of chain conformations is possible for 
DNA molecules \cite{mai99,mai00,mai01},
nanorope polymer chains \cite{wang04} or
brushlike polymers \cite{dobrynin07} 
adsorbed on strongly attractive surfaces or confined in thin films 
by means of fluorescence microscopy \cite{mai99,mai00} or 
atomic force microscopy \cite{wang04,dobrynin07}. 
The experimental verification of the various conformational properties
discussed in this paper, such the one described by Eq.~(\ref{eq_compact}),
is thus conceivable for these systems \cite{dobrynin07}.
%

%
%
%
Compactness does obviously {\em not} imply Gaussian chain statistics \cite{DegennesBook}
(as incorrectly stated, e.g., in \cite{wang04}) 
nor does segregation of chains and subchains impose disk-like shapes 
minimizing the average perimeter length $L(s)$ of chains and subchains \cite{ANS03}. 
The contour boundaries are in fact found to be highly irregular 
as is revealed by the snapshots presented in Fig.~\ref{fig_snapshot} and Fig.~\ref{fig_snapblob}. 
Elaborating a short communication made recently \cite{MKA09} we present here theoretical arguments and 
molecular dynamics simulations demonstrating that the contours are in fact {\em fractal},
scaling as
\begin{equation}
L(s) \sim R(s)^{\dc} \sim s^{1- \nu \Theta_2},
\label{eq_keyLs}
\end{equation}
with $\dc = d- \Theta_2 = 5/4 > 1$ being the fractal line dimension \cite{Mandelbrot}. 
Our work is based on the pioneering work by Duplantier who predicted a contact exponent $\Theta_2 = 3/4$ \cite{dupl} 
and a more recent paper by Semenov and one of us (A.J.) \cite{ANS03}.
In contrast to many other possibilities to characterize numerically the compact chain 
conformations the perimeter length is of interest since it can be related
to the intrachain structure factor $F(q)$ with $\vec{q}$ being the wave vector. 
It is thus accessible experimentally, at least in principle, by means of 
small-angle neutron scattering experiments to all polymers
which can be appropriately labeled \cite{BenoitBook}. 
Specifically, it will be demonstrated that due to the generalized Porod scattering 
of {\em compact} objects \cite{BenoitBook,bale84,bray88} 
the structure factor of dense 2D polymers should scale 
in the intermediate wave vector regime as 
\begin{equation}
F(q)  \approx N/(q R(N))^{d+\Theta_2} \sim N^{-3/8} q^{-11/4}
\label{eq_keyFq}
\end{equation}
for sufficiently long chains and
{\em not} as $F(q) \sim N^0 q^{-2}$ as numerous authors have assumed 
\cite{carmesin90,mai99,pakula02,yethiraj03}.
%

%
The present paper is organized as follows.
In Sec.~\ref{sec_algo} we recall the computational model used for this study.
Our mainly numerical results are presented in Sec.~\ref{sec_results}.
We confirm first the compactness of the chain conformations by
considering the typical size of chains and subchains (Sec.~\ref{sub_Rs}).
Corrections to this leading power-law behavior due to chain-end effects caused 
by the confinement of the chains will be analyzed in Sec.~\ref{sub_Rscorrect}. 
That $\nu=1/2$ does {\em not} imply Gaussian chain statistics will be emphasized by 
the scaling analysis of various intrachain properties such as 
the histograms of inner chain distances $G_i(r)$ in Sec.~\ref{sub_theta},
the bond-bond correlation functions $P_1(s)$ and $P_2(s)$ in Sec.~\ref{sub_bond}
or the single chain structure factor $F(q)$ in Sec.~\ref{sub_Fq}.
Two (related) scaling arguments will be given in Sec.~\ref{sub_Ls} and at the end of Sec.~\ref{sub_Fq}, 
demonstrating the scaling of the perimeter length $L(s)$, Eq.~(\ref{eq_keyLs}).
The analytic calculation of the structure factor $F(q)$ for 2D melts is relegated to the Appendix.
A discussion of possible consequences of the observed static properties for the dynamics of
dense 2D solutions and melts concludes the paper in Sec.~\ref{sec_concl}.
%

%
\section{Computational details}
\label{sec_algo}

Our numerical results are obtained by standard molecular dynamics simulations of monodisperse 
linear chains at high densities. The coarse-grained polymer model Hamiltonian is essentially identical 
to the standard Kremer-Grest (KG) bead-spring model \cite{KG86,KG90} which has been used in numerous 
simulation studies of diverse problems in polymer physics \cite{KG86,KG90,duen,NIC}.
The non-bonded excluded volume interactions are represented by a purely repulsive Lennard-Jones potential
\cite{FrenkelSmitBook}
\begin{equation}
U_{\text{nb}}(r) = 4 \epsilon \left[ (\sigma/r)^{12} - (\sigma/r)^6 \right] + \epsilon 
\mbox{ for } r/\sigma \le 2^{1/6} 
\label{eq_algoLJ}
\end{equation}
and $U_{\text{nb}}(r) = 0$ elsewhere. 
The Lennard-Jones potential does {\em not} act between adjacent monomers of a chain 
which are topologically connected by a simple harmonic spring potential
\begin{equation}
U_{\text{b}}(r) = k_{\text{b}} (r - l_{\text{b}})^2
\label{eq_algoSPRING}
\end{equation}
with a spring constant $k_{\text{b}} = 338 \epsilon$ and a bond reference length 
$l_{\text{b}} = 0.967 \sigma$. Both constants have been calibrated to the 
``finite extendible nonlinear elastic" (FENE) springs of the original KG model. 
Lennard-Jones units are used throughout this paper ($\epsilon=\sigma=1$).
The classical equations of motion of the multichain system are solved via the Velocity-Verlet 
algorithm at constant temperature using a Langevin thermostat with friction constant $\gamma=0.5$ 
\cite{FrenkelSmitBook}.
We focus in this presentation on melts of density $\rho=7/8=0.875$ at temperature $T=1$. 
Due to the excluded volume potential monomer overlap is strongly reduced 
as may be seen from the pair correlation function $g(r)$ shown in Fig.~\ref{fig_gr}.
The bonding potential, Eq.~(\ref{eq_algoSPRING}), prevents the long range correlations 
which would otherwise occur at such a high density for a 2D system of monodisperse 
Lennard-Jones beads \cite{ChaikinBook,FrenkelSmitBook}. 
As shown in the inset of Fig.~\ref{fig_gr} we have in fact a dense {\em liquid} and the oscillations 
of $g(r)-1$ decay rapidly with an exponential cut-off.
The parameters of the model Hamiltonian and the chosen density and temperature makes chain 
intersections impossible, as can be seen from the snapshot of ``chain 1" presented in Fig.~\ref{fig_snapshot}.
We simulate thus ``self-avoiding walks" in the sense of 
the first model class discussed in Ref.~\cite{ANS03}.

Monodisperse systems with chain lengths $N$ ranging between $N=32$ up to $N=2048$
have been sampled using periodic square boxes containing either $98304$ or $196608$ monomers.
The larger box of linear length $474.02$ corresponds to $96$ chains of length $N=2048$.
Some conformational properties discussed below are summarized in the Table.
Except the systems with $N=2048$, all chains have diffused over at least 10 times 
their radius of gyration $\Rgyr(N)$ providing thus sufficiently good statistics.
Note that our largest chain is about an order of magnitude larger than the largest
chains used in previous computational studies of dense 2D melts:
$N = 59$ by Baumg\"artner in 1982 \cite{baumgartner82},
$N = 100$ by Carmesin and Kremer in their seminal work in 1990 \cite{carmesin90}, 
$N = 100$ by Nelson {\em et al.} in 1997 \cite{rutledge97},
$N = 32$ by Polanowski and Pakula in 2002 \cite{pakula02},
$N = 60$ by Balabaev {\em et al.} in 2002 \cite{balabaev02},
$N = 256$ by Yethiraj in 2003 \cite{yethiraj03} 
and
$N = 256$ by Cavallo {\em et al.} in 2005 \cite{cavallo05,CMWJB05}.
The presented data was obtained on IBM power 6 with the LAMMPS Version 21May2008 \cite{LAMMPS}.
It is part of a broader study where we have systematically varied density, system size
and friction coefficient to confirm the robustness of theory and simulation with respect to these parameters.
%
Since the chain length is computationally the limiting factor fixing the number of ``blobs" 
\cite{DegennesBook} at a given density we present data at the largest density,
i.e. the largest number of blobs, where we have been able to equilibrate chains of $N=2048$.

\section{Numerical results}
\label{sec_results}

\subsection{Chain and subchain size: Compactness}
\label{sub_Rs}
%

%
Figure~\ref{fig_Rs} confirms that 2D chains are indeed compact as stated in Eq.~(\ref{eq_compactN})
and Eq.~(\ref{eq_compact}) and as shown in Fig.~\ref{fig_snapshot} for chains and in Fig.~\ref{fig_snapblob} 
for subchains of arc-length $s$.
The typical size $R(s)$ of subchains is characterized by either the root-mean-square 
end-to-end distance $\Rend(s)$ or the radius of gyration $\Rgyr(s)$ \cite{DoiEdwardsBook,WBM07}.
As indicated in the sketch we consider a subchain between two monomers $n$ and $m=n+s-1$ 
and average over all pairs $(n,m)$ possible in a chain of length $N$ following \cite{WMBJOMMS04,Auhl03,WBM07}.
Averaging only over subchains at the curvilinear chain center ($n,m \approx N/2$) slightly reduces 
chain end effects, however the difference is negligible for the larger chains, $N \ge 1024$, we focus on.
The limit $s=N$ corresponds obviously to the standard end-to-end vector $\Rend(N)$ and 
gyration radius $\Rgyr(N)$ of the total chain.
Open symbols refer to subchains of length $s \le N$ with $N=1024$ (squares) and $N=2048$ (spheres), 
stars to total chain properties ($s=N$).
In agreement with various numerical \cite{baumgartner82,carmesin90,rutledge97,pakula02,balabaev02,yethiraj03,cavallo05} and
experimental studies \cite{mai99,dobrynin07} the presented data confirms that the chains are compact, 
i.e. $\nu=1/d$ (thin lines), and this on all length scales with $1 \ll s \le N$.

The segregation of the chains may also be shown by computing the average number
$\nchain$ of chains in contact with a reference chain, i.e. having at
least one monomer closer than a distance $a \approx 2$ to a monomer of the
reference chain (Sec.~\ref{sub_Ls}). Depending weakly on $a$, we find $\nchain \approx 6$,
as one may expect for 2D colloids and in agreement with Fig.~\ref{fig_snapshot}.
At variance to any open non-segregated polymer-like structure $\nchain$ does thus
{\em not} increase with chain length $N$. An alternative way to confirm this statement
is to count the centers of mass around the reference chain's center of mass
by integrating the center-of-mass pair-correlation function $\gcm(r)$,
a structureless function without oscillations (not shown). One verifies that
6 chains are found in a shell of about $2 \Rgyr(N)$ and this irrespective of $N$.

\subsection{Segment size distributions and contact exponents}
\label{sub_theta}
Being characterized by the same Flory exponent $\nu$ as their three dimensional (3D) counterparts does 
by no means imply that 2D melts are Gaussian \cite{dupl,ANS03}.  This can be directly seen, e.g., 
from the different probability distributions of the intrachain vectors $\vec{r} = \vec{r}_m - \vec{r}_n$ 
presented in Fig.~\ref{fig_theta}.
To simplify the plot we have focused on the two longest chains $N=1024$ and $N=2048$ we have simulated.
As illustrated in panel (a), 
\begin{itemize}
\item $G_0(r,s=N)$ characterizes the distribution of the total chain end-to-end vector ($n=1$, $m=N$), 
\item $G_1(r,s=N/2)$ the distance between a chain end and a monomer in the middle of the chain ($n=1$, $m=N/2$), 
\item $G_2(r,s=N/2)$ the distribution of an inner segment vector between the monomers $n=N/4$ and $m=3N/4$,
\item while the ``segmental size distribution" $\Ge(r,s)$ \cite{WBM07} {\em averages} over all pairs of monomers 
$(n,m=n+s-1)$ for $s \le N$.
\end{itemize}
The second index $s$ indicated in $G_i(r,s)$ characterizes the length of the subchain 
between the two monomers $n$ and $m$. 
As shown in panel (b), all data for different $N$ and $s$ collapse on three distinct master curves 
if the axes are made dimensionless using the second moment $R_i^2$ of the respective distribution.
The only relevant length scale is thus the typical size of the subchain itself. 
The distributions are all non-monotonous and are thus {\em qualitatively} different from the Gaussian
(thin line) expected for uncorrelated ideal chains. Confirming Duplantier's predictions \cite{dupl}
we find 
\begin{equation}
R_i^d G_i(r,s) = x^{\Theta_i} f_i(x)
\label{eq_theta}
\end{equation}
with $x=r/R_i$ being the scaling variable and the contact exponents $\Theta_0=3/8$,
$\Theta_1=1/2$ and $\Theta_2=3/4$ (dashed lines) describing the small-$x$ limit where 
the universal functions $f_i(x)$ become constant. 
Especially the largest of these exponents, $\Theta_2$, is clearly visible. The contact 
probability for two monomers of a chain in a 2D melt is 
thus strongly suppressed compared to ideal chain statistics ($\Theta_0=\Theta_1=\Theta_2=0$).
The rescaled distributions show exponential cut-offs for large distances. 
The Redner-des Cloizeaux formula \cite{DescloizBook} is a useful interpolating
formula which supposes that 
\begin{equation}
f_i(x) = c_i \exp(-k_i x^2).
\label{eq_Redner}
\end{equation}
The constants $k_i=1+\Theta_i/2$ and $c_i = k_i^{k_i}/\pi\Gamma(k_i)$ 
with $\Gamma(z)$ being the Gamma function \cite{abramowitz} are imposed
by the normalization and the second moment of the distributions \cite{ever95}. 
This formula is by no means rigorous but yields reasonable parameter free fits 
as it is shown by the solid line for $f_2(x)$. 

Obviously,  $\Ge(r,s) \approx G_0(r,N)$ for very large subchains $s \rightarrow N$ (not shown).
As can be seen, the rescaled distributions $\Ge(r,s)$ and $G_2(r,N/2)$ become identical if 
$1 \ll s \ll N$.
It is for this reason that the exponent $\Theta_2$ is central for asymptotically long chains
as will become obvious below in Sec.~\ref{sub_Ls} and Sec.~\ref{sub_Fq}.
The two exponents $\Theta_0$ and $\Theta_1$ are only relevant if the measured property 
specifically highlightes chain end effects as in the example given in the next subsection.

\subsection{Chain and subchain size: Corrections to asymptotic scaling}
\label{sub_Rscorrect}

The log-log representation chosen in Fig.~\ref{fig_Rs} masks deliberately small corrections to the 
leading power law due to chain end effects which exist in 2D as they do in 3D melts \cite{WBM07}. 
These are revealed in Fig.~\ref{fig_RsB} presenting $\Rend(s)/s^{1/2}$ and $\Rgyr(s)/s^{1/2}$ {\em vs.} $s$ 
and $\Rend(N)/N^{1/2}$ and $\Rgyr(N)/N^{1/2}$ {\em vs.} $N$
using log-linear coordinates and the same symbols as in Fig.~\ref{fig_Rs}.
The reduced radius of gyration (bottom data) becomes in fact rapidly constant and chain length independent.
As emphasized by the bottom horizontal line we find 
\begin{equation}
\bg \approx  \Rgyr(s)/s^{1/2} \approx \Rgyr(N)/N^{1/2}
\mbox{ for } s,N \gg 100
\label{eq_bg}
\end{equation}
with $\bg=0.65$ for $\rho=7/8$.
Interestingly, we observe {\em non-monotonous} behavior for $\Rend(s)/s^{1/2}$ with a decay for $s > N/2$.
Due to this decay $\Rend(N)/N^{1/2}$ (stars) is systematically {\em below} the corresponding 
internal chain distance $\Rend(s)/s^{1/2}$.
If fitted from the chain end-to-end distances one obtains an effective segment size \cite{DoiEdwardsBook}
\begin{equation}
\bestar \approx \Rend(N)/N^{1/2} \approx 1.5 \mbox{ for } 100 \ll N,
\label{eq_bestar}
\end{equation}
as shown by the dashed line.
(The index $\star$ indicates that we refer to the total chain.) 
This value corresponds to the ratio $\Rend^2(N)/\Rgyr^2(N) \approx 5.3 < 6$ given in the Table.
It confirms similar observations made in previous simulations using much shorter chains 
\cite{baumgartner82,carmesin90,rutledge97,yethiraj03}.
If on the other hand the effective segment size $\be$ is obtained from the 
internal distances this yields 
\begin{equation}
\be \approx \Rend(s)/s^{1/2} \approx 1.6 \mbox{ for } 100 \ll s \ll N,
\label{eq_beRs}
\end{equation}
as indicated by the top solid line. 
This value is consistent with a ratio $\Rend^2(s)/\Rgyr^2(s)=(\be/\bg)^2 \approx 6$ 
as one expects in any dimension due to \cite{DoiEdwardsBook}
\begin{equation}
\Rgyr^2(s) = \frac{1}{s^2} \int_0^s d\sprim (s-\sprim) \Rend^2(\sprim) = \frac{1}{6} \be^2 s
\label{eq_Rend2Rgyr}
\end{equation} 
where we have {\em assumed} $\Rend^2(\sprim) = \be^2 \sprim$ for {\em all} $\sprim$ up to $\sprim=s$.
Since this assumption breaks down for $s \to N$ (as seen in Fig.~\ref{fig_RsB}) 
Eq.~(\ref{eq_Rend2Rgyr}) is not in conflict with Eq.~(\ref{eq_bestar}).
Note that the integral over $\sprim$ in Eq.~(\ref{eq_Rend2Rgyr}) is dominated by
subchains with $\sprim \approx s/2$ and that thus large subchains of order $\sprim \approx N$
are less relevant for the gyration radius of the total chain $\Rgyr(s=N)$. 
Hence, the non-monotonous behavior observed for $\Rend(s)$ should be barely detectible
for the radius of gyration in agreement with Eq.~(\ref{eq_bg}). 

That a naive fit of $\be$ from the total chain end-to-end distance $\Rend(N)$ leads to a
systematic {\em underestimation} of the effective segment size of asymptotically long chains 
is a well-known fact for 3D melts \cite{WBM07}. However, both estimations of $\be$ 
merge for 3D melts if sufficiently long chains are computed 
(as may be seen from Eq.~(16) and Fig.~4 of Ref.~\cite{WBM07}).
Apparently, this is not the case in 2D since 
if $\Rend(s)$ or $\Rend^2(s)/s$ are plotted as a function of $x=s/N$ 
a nice scaling collapse of the data is obtained for large $x$ and for $N \ge 256$, i.e.
\begin{equation}
\Rend^2(s) \approx \be^2 s^{2\nu} \fRsN(x) \mbox{ with } 
\fRsN(x) = \left\{
\begin{array}{ll}
1 & \mbox{ if $x \ll 1$} \\
(\bestar/\be)^2 & \mbox{ if $x \to 1$}.
\end{array}
\right.
\label{eq_RsNscaling}
\end{equation}
(Since a very similar scaling plot is presented in the inset of Fig.~\ref{fig_P1} this figure is not given.)
Hence, chain end effects may {\em not} scale away with $N \to \infty$ as they do in 3D.
A simple qualitative explanation for Eq.~(\ref{eq_RsNscaling})
is in fact readily given by considering an ideal chain of bond length $\be$ squeezed into a more 
or less spherical container of size $R_c \sim N^{1/2}$. For $s \ll N$ it is unlikely that the 
subchain interacts with the container walls and $\Rend(s) \approx \be s^{1/2}$. With $s \to N$ 
the chain will feel increasingly the confinement reflecting it back from the walls into the center 
of the container reducing thus the effective segment length $\bestar$ associated with the 
chain end-to-end distance.
%
%
Since the scaling function $\fRsN(x)$ must be universal, it should be possible to express the ratio 
$(\bestar/\be)^2$ 
--- and thus the ratio $\Rend^2(N)/\Rgyr^2(N) = (\bestar/\bg)^2$ ---
in terms of the dimension $d$ and universal compact exponents $\Theta_0$, $\Theta_1$ and $\Theta_2$. 
The two statistical segment sizes $\be$ and $\bestar$ should thus be related.
At present we are still lacking a solid theoretical proof for the latter conjection. 
%

\subsection{Intrachain orientational correlations}
\label{sub_bond}
%

%
Let $\vec{e}_i$ denote the normalized tangent vector connecting the monomers $i$ and $i+1$ of a chain.
The bond-bond correlation function $P_1(s) = \langle \vec{e}_n \cdot \vec{e}_m \rangle$ 
has been shown to be of particular interest for characterizing the deviations from 
Gaussianity in 3D polymer melts \cite{WMBJOMMS04,WBM07}. 
(As above we average over all pairs of monomers $(n,m)$ with $m=n+s-1$.)
The reason for this is that \cite{WMBJOMMS04} 
\begin{equation}
P_1(s) \sim - \frac{\partial^2 \Rend^2(s)}{\partial s^2}
\label{eq_P1Rs}
\end{equation}
and that thus small deviations from the asymptotic exponent $2\nu=1$ are emphasized \cite{WBM07}. 
$P_1(s)$ is presented in Fig.~\ref{fig_P1} for different chain lengths.
Apparently, the bond pairs are aligned only for small arc-lengths with $s < 10$.
For larger $s$ the bonds are {\em anti-correlated} with two characteristic peaks.
The first anti-correlation peak visible in panel (a) is due to the local backfolding
of the chain contour which can be directly seen from chain 1 drawn in Fig.~\ref{fig_snapshot}.
Note that the chain length dependence disappears for $N > 256$. 
The second anti-correlation peak is shown in panel (b) where $-P_1(s) N$ is plotted as function of $x=s/N$.
Using Eq.~(\ref{eq_P1Rs}) this corresponds exactly to the scaling expected from Eq.~(\ref{eq_RsNscaling})
with an associated universal function scaling as $-\partial_x^2 ( x \fRsN(x) )$.
In agreement with the qualitative explanation mentioned at the end of Sec.~\ref{sub_Rs},
the peak at $s \approx N/2$ can be attributed to the confinement of the chain which causes long
segments to be reflected back, i.e. $P_1(s)$ must be anti-correlated for large $x$. 
(Data for $N=2048$ is not presented here due to its insufficient statistics.)
We stress finally that altogether this is a rather small effect and essentially 
$P_1(s) \approx \partial^2_s \Rend^2(s) \approx 0$ for $s \gg 10^2$ and $N \gg 10^3$ as already 
obvious from panel (a). Hence, the first Legendre polynomial 
confirms that to leading order $2 \nu \approx 1$ for sufficiently large chains and segments.
 
%
Conceptually more important for the present study is the fact that the second Legendre polynomial 
$P_2(s) = \la \left(\vec{e}_n \cdot \vec{e}_m\right)^2 \ra -1/2$ given in the main panel of Fig.~\ref{fig_P2}
reveals a clear power law behavior over two orders of magnitude in $s$  (dashed line).
This power law is due to 
{\em (a)} the return probability $p_r(s)$ after $s$ steps and
{\em (b)} the ``nematic alignment" of two near-by bonds.
The alignment of bonds is investigated in the inset of Fig.~\ref{fig_P2} where the 
second Legendre polynomial is plotted as a function of the distance $r$ between 
the mid-points of both bonds. Averages are taken over all intrachain bond pairs 
with $[r,r+\delta r]$ using a bin of width $\delta r = 0.01$. Since $P_2(r)$
becomes rapidly chain length independent we only indicate data for $N=2048$.
The vertical axis is rescaled with the phase volume $2\pi r$.
As can be seen, the orientational correlations oscillate with $r$
and this with a rapidly decaying amplitude. 
These oscillations are related to the oscillations of the pair correlation function $g(r)$
shown in Fig.~\ref{fig_gr} and reflect the local packing and wrapping of chains
composed of discrete spherical beads.
Due to both the oscillations and the decay
only bond pairs at $r \approx 1$ matter if we compute $P_2(s)$, i.e. if we sum over 
all distances $r$ at a fixed curvilinear distance $s$.
Following Eq.~(\ref{eq_theta}) one thus expects
\begin{equation}
P_2(s) \approx p_r(s) \equiv \lim_{r\rightarrow 0} \Ge(r,s) \sim 1/s^{1+\nu\Theta_2}=1/s^{11/8}
\label{eq_preturn}
\end{equation}
for $1 \ll s \ll N$.
The agreement of the data with this power law is excellent and provides, hence, 
an independent confirmation of the contact exponent $\Theta_2=3/4$.

\subsection{Chain and subchain shape}
\label{sub_asph}
As obvious from Fig.~\ref{fig_snapshot} and Fig.~\ref{fig_snapblob} the conformations
of chains and subchains are neither perfectly spherical nor extremely elongated. 
Having discussed above the chain size we address now the chain shape as characterized 
by the average {\em aspherity} of the gyration tensor. 
The gyration tensor $\Mmat$ of a subchain between the monomers $n$ and $m=n+s$ is given by
\begin{equation}
M_{\alpha \beta} = \frac{1}{s} \sum_{i=n}^m 
(r_{i,\alpha} - R_{\text{cm},\alpha}) (r_{i,\beta} - R_{\text{cm},\beta})
\label{eq_inertia}
\end{equation}
with $R_{\text{cm},\alpha}$ being the $\alpha$-component of the subchains's center of mass.
We remind that the radius of gyration $\Rgyr^2(s)$ discussed in Sec.~\ref{sub_Rs}
is given by the trace $\trace(\Mmat) = M_{xx} + M_{yy} = \lambda_1 + \lambda_2$ 
averaged over all subchains and chains with eigenvalues $\lambda_1$ and $\lambda_2$ 
obtained from 
\begin{equation}
\lambda_{1,2} = \frac{1}{2} \left( \trace(\Mmat) \pm \sqrt{ \trace(\Mmat)^2 - 4 \deter(\Mmat)} \right).
\label{eq_lambda}
\end{equation}
The ratio of the mean eigenvalues $\la \lambda_1\ra/\la \lambda_2\ra$ for $s=N$ is given in the Table.
Decreasing slightly with $N$ this ``aspect ratio" approaches 
\begin{equation}
\la \lambda_1 \ra : \la \lambda_2 \ra \approx 4.5 : 1
\label{eq_aspectratio}
\end{equation}
for our longest chains
which corresponds to a reduced principal eigenvalue $\la \lambda_1 \ra/\Rgyr^2 \approx 0.8$.
It should be noted that Gaussian chains and dilute good solvent chains in 2D are characterized
by an aspect ratio $\la \lambda_1\ra/\la \lambda_2\ra \approx 5.2$ and $6.7$, 
respectively \cite{Bishop86}. Our chains are thus clearly less elongated.
The asphericity of the inertia tensor of 2D objects may be further characterized by 
computing the moments \cite{rudnick86,nelson86,Bishop86,mai01}
\begin{equation}
\Delta_1(s) = \frac{\la \lambda_1-\lambda_2 \ra}{\la \lambda_1+\lambda_2 \ra} 
\ , \
\Delta_2(s) = \frac{\la (\lambda_1-\lambda_2)^2 \ra}{\la (\lambda_1+\lambda_2)^2 \ra} 
\label{eq_Deltadef}
\end{equation}
which are plotted in Fig.~\ref{fig_asph} for subchains ($s \le N$) and total chains ($s=N$) 
using the same symbols as in previous plots.
$\Delta_1 = 2 \la \lambda_1 \ra/\Rgyr^2 -1$ describes the mean ellipticity
and $\Delta_2$ the normalized variance of $\lambda_1$ and $\lambda_2$ \cite{rudnick86,nelson86}. 
Obviously, $\Delta_1=\Delta_2=1$ for rods and $\Delta_1=\Delta_2=0$ for spheres.
Note that taking the first and the second moments of the ellipticity 
$(\lambda_1 - \lambda_2)/(\lambda_1+\lambda_2)$ of {\em each} subchain
yields qualitatively similar results (not shown).
As one expects, both moments do not depend on whether a chain or a subchain is considered. 
In agreement with Yethiraj \cite{yethiraj03} they decrease weakly with $N$ and $s$. 
Unfortunately, it is difficult to determine precisely the plateau values one expects for 
asymptotically large chains and subchains, and the horizontal lines with 
\begin{equation}
\Delta_1 = 0.63 \mbox{ and } \Delta_2 = 0.51
\label{eq_Deltavalues}
\end{equation}
are merely guides to the eye.
Note that Yethiraj \cite{yethiraj03} indicates $\Delta_2 = 0.52$ for $N=256$.
Considering that the latter value has been obtained at a slightly smaller monomer volume fraction
both $\Delta_2$-values are compatible.
We remind that in two dimensions $\Delta_1 \approx 0.68$ and 
$\Delta_2 = 2 (d+2)/(5d+4) \approx 0.57$ 
for Gaussian chains \cite{rudnick86,Bishop86} and 
$\Delta_1 \approx 0.74$ and $\Delta_2 \approx0.62 - 0.64$ 
for dilute good solvent chains according to Refs.~\cite{nelson86,Bishop86,mai01,yethiraj03}. 
These values are definitely {\em larger} than our respective estimates, Eq.~(\ref{eq_Deltavalues}), 
and segregated chains in 2D melts are thus clearly more axisymmetric.

The above analysis has been motivated by recent experimental work on the conformational 
properties of {\em dilute} DNA molecules investigated using fluorescence microscopy \cite{mai01}.
A similar characterization of the chain shapes at higher semidilute densities
appears therefore feasible, at least in principle.

\subsection{The perimeter length}
\label{sub_Ls}

\begin{figure}[tb]
\includegraphics*[width=12.0cm]{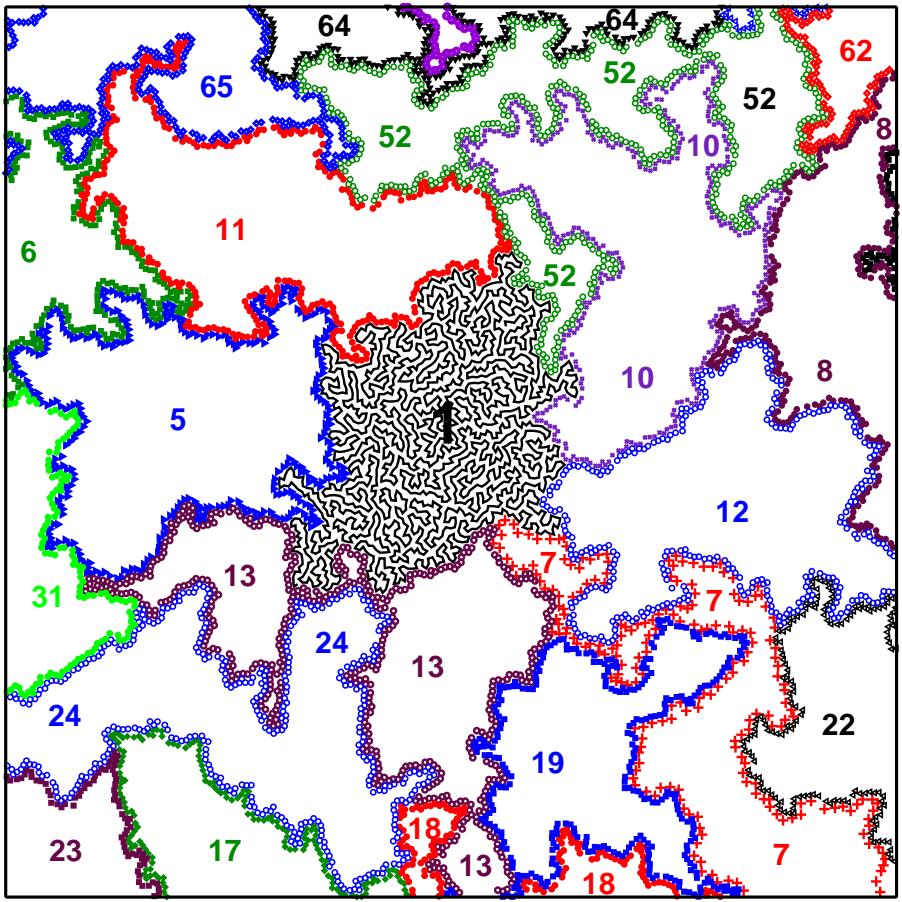}
\caption{Snapshot of a 2D polymer melt with chain length $N=2048$ and monomer number density $\rho=7/8$. 
Due to the excluded volume interactions the monomers do barely overlap and chain intersections 
are strictly forbidden.
One chain in the middle is fully drawn while for the other chains only the perimeter
monomers interacting with other chains are indicated. A perimeter monomer is defined here
as having at least one monomer {\em not} belonging to the same chain within a distance $r \le a =1.2$.
Numbers refer to an arbitrary chain index used for computational purposes. 
The chains are 
compact, i.e., they fill space densely, and interact typically with about $6$ other chains. 
However, compactness does apparently not imply a disk-like shape 
which would minimize the perimeter length $L(N)$. 
Quantitative analysis reveals that the contours are fractal with a fractal line dimension $\dc =5/4$. 
\label{fig_snapshot}
}
\end{figure}

\begin{figure}[tb]
\includegraphics*[width=12.0cm]{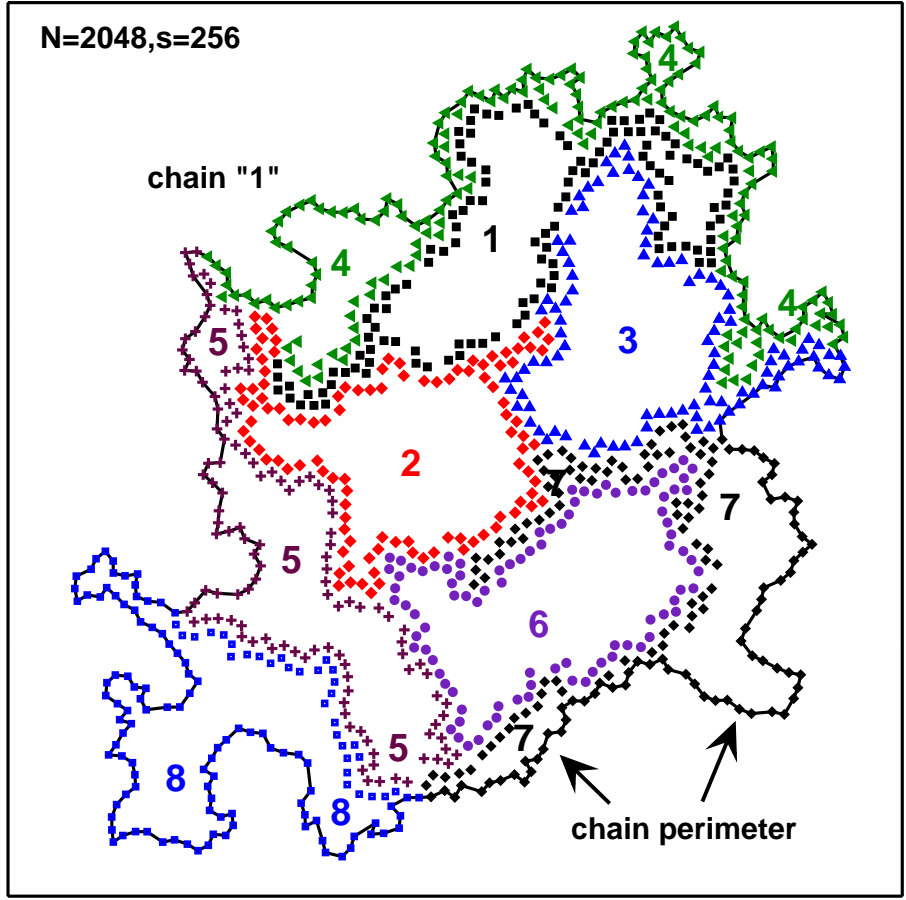}
\caption{Self-similarity of compactness and perimeter fractality on all scales shown for the chain ``1" 
from the snapshot Fig.~\ref{fig_snapshot}. The solid line indicates the perimeter of this chain
with respect to monomers of other chains. We consider 8 consecutive subchains of length $s=256$ and 
compute their respective perimeter monomers being close to monomers from other chains or subchains.
The subchains are compact and of irregular shape, just as the total chains ($s=N$) presented in Fig.~\ref{fig_snapshot}.  
\label{fig_snapblob}
}
\end{figure}

\begin{figure}[tb]
\includegraphics*[width=12.0cm]{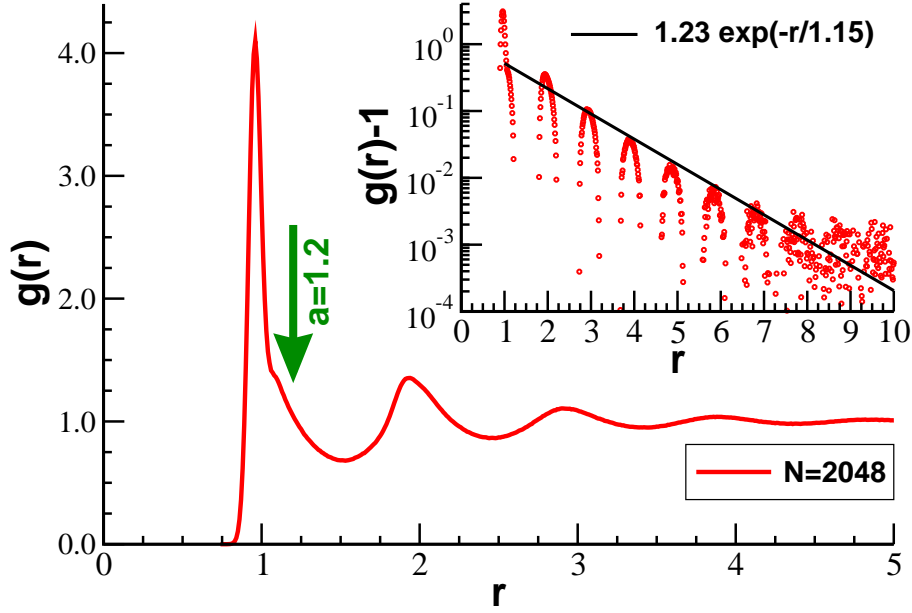}
\caption{Radial pair correlation function $g(r)$ between all monomers for chains of length $N=2048$
at monomer number density $\rho=7/8$. Note that $g(r) \to 1$ for large monomer distances $r$.
The correlation function oscillates strongly with a period given by the bead diameter $\sigma=1$. 
As shown in the inset, the amplitude of these oscillations decays exponentially as expected
for a dense liquid without long-range positional order \cite{ChaikinBook}.
The arrow indicates the position of the reference length $a=1.2$ used 
for identifying a monomer as being in contact with a monomer from
another chain or subchain.
\label{fig_gr}
}
\end{figure}

\begin{figure}[tb]
\includegraphics*[width=12.0cm]{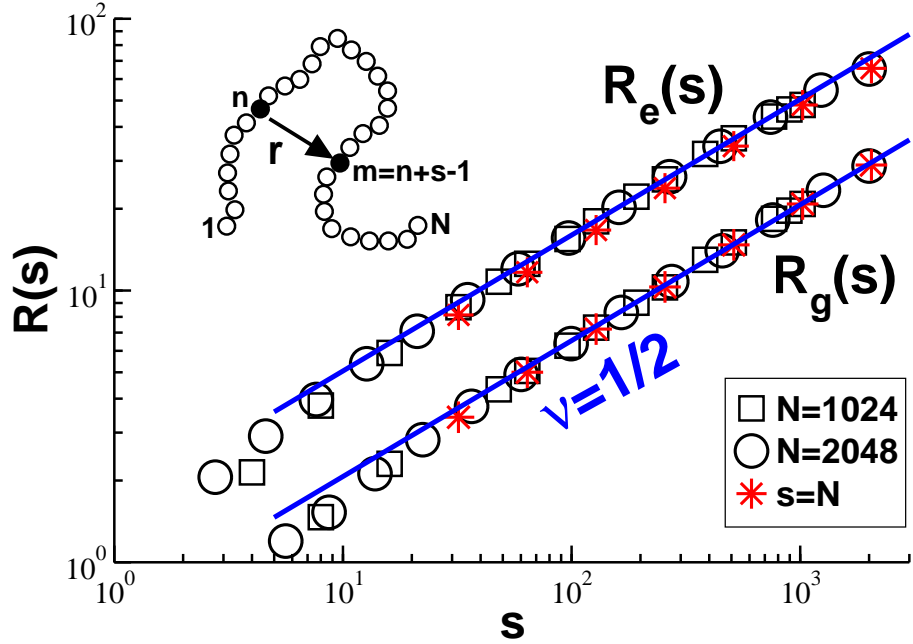}
\caption{End-to-end distance $\Rend(s)  = \langle \vec{r}^2 \rangle^{1/2}$ (top data) and
radius of gyration $\Rgyr(s)$ (bottom data) of ``subchains" or ``chain segments" \cite{WBM07}
containing $s=m-n+1$ monomers as indicated by the sketch on the left hand side. 
The data is averaged over all $s$-subchains possible in a chain of length $N$.
Open symbols refer to subchains with $s \le N$ for $N=1024$ (squares) and $N=2048$ (spheres),
stars to overall chain properties ($s=N$).
The indicated power-law slopes confirm Eq.~(\ref{eq_compact}), i.e. the chains are compact on all scales.
\label{fig_Rs}
}
\end{figure}

\begin{figure}[tb]
\includegraphics*[width=10.0cm]{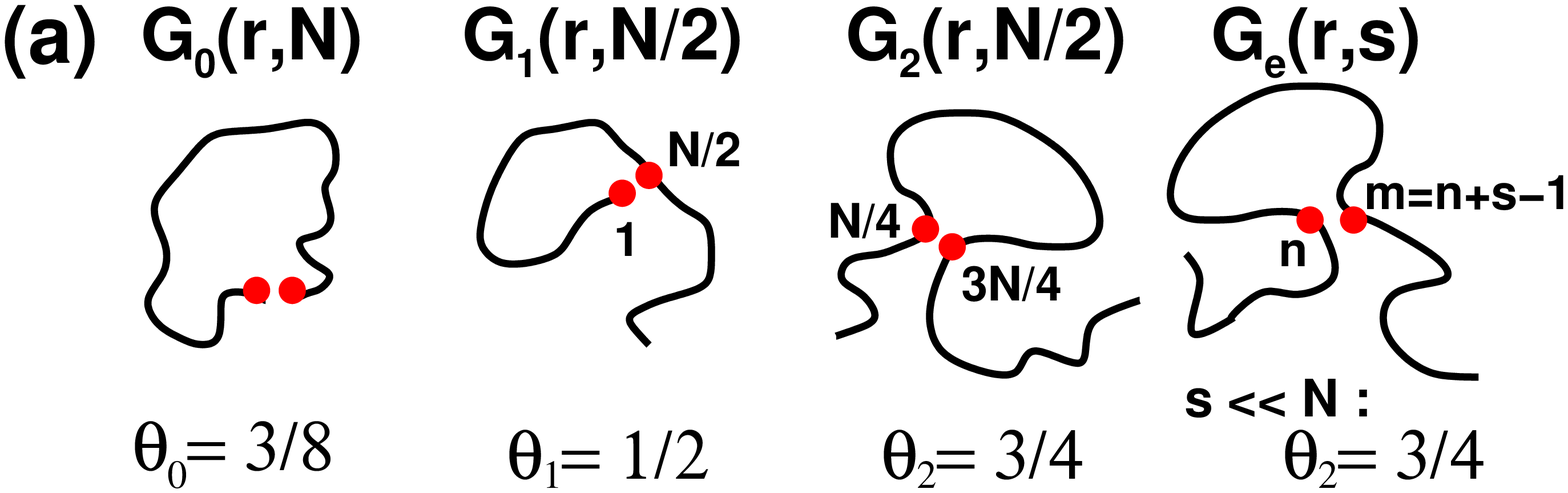}

\vspace*{0.5cm}
\includegraphics*[width=12.0cm]{fig_theta}
\caption{{\bf (a)} Sketch of the intrachain probability distributions $G_i(r,s)$ 
described in the main text and the contact exponents $\Theta_i$ predicted by 
Duplantier \cite{dupl}.
{\bf (b)} Scaling plots are presented for $G_0(r,N)$, $G_1(r,N/2)$ and $G_2(r,N/2)$ 
for $N=1024$ (squares) and $N=2048$ (spheres)
and for $\Ge(r,s)$ for $N=1024 \gg s$ with $s=256$ (triangles) and $s=512$ (diamonds).
All data for different $N$ and $s$ collapse on the respective universal master curves 
if $y = R_i^2 G_i(r,s)$ is plotted {\em vs.} the reduced distance $x=r/R_i$ with $R_i^2$ 
being the second moment of the corresponding distribution.
The thin line indicates the Gaussian distribution $y = \exp(-x^2)/\pi$ expected for ideal chains in 2D.
The power laws $y \approx x^{\Theta_i}$ observed for $x \ll 1$ (dashed lines) confirm Duplantier's prediction.
The solid line at the bottom shows the Redner-des Cloizeaux formula, Eq.~(\ref{eq_Redner}), 
for $G_2(r,s=N/2) \approx \Ge(r,s)$.
\label{fig_theta}
}
\end{figure}

\begin{figure}[tb]
\includegraphics*[width=12.0cm]{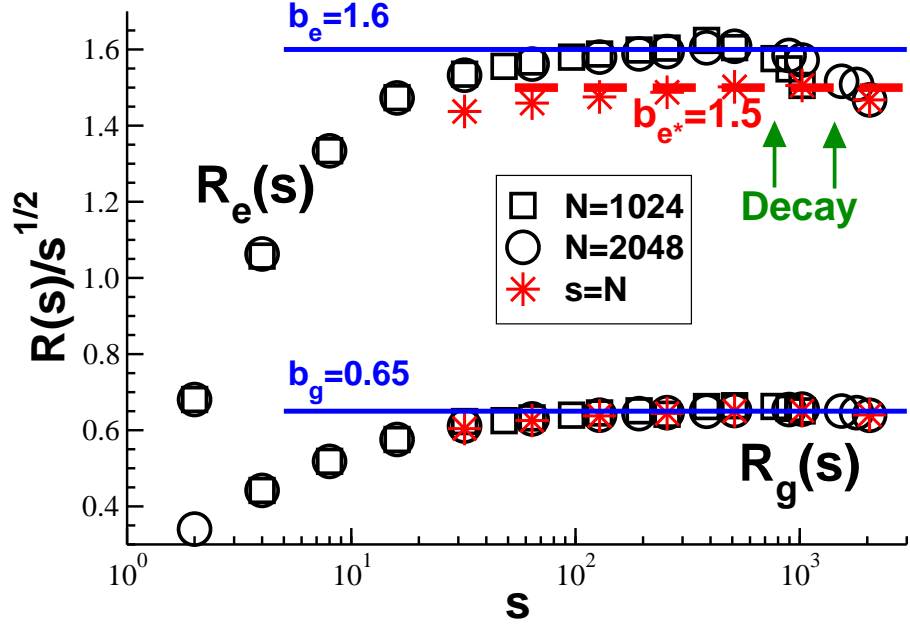}
\caption{Reduced typical subchain size $R(s)/s^{1/2}$ as characterized by 
the root-mean-square end-to-end distance $\Rend(s)$ (top data) 
and the radius of gyration $\Rgyr(s)$ (bottom data) using the same symbols as in Fig.~\ref{fig_Rs}.
Interestingly, $\Rend(s)/s^{1/2}$ decays for $s > N/2$ due to chain-end effects.
The solid horizontal lines indicate the effective segment sizes $\be=1.6$ and $\bg=0.65$ 
obtained for $100 \ll s \ll N$ from $\Rend(s)/s^{1/2}$ and $\Rgyr(s)/s^{1/2}$, respectively. 
The dashed line corresponds to an apparent effective segment size $\bestar=1.5$ from the 
end-to-end distances $\Rend(N)$ of our longest chains.
\label{fig_RsB}
}
\end{figure}

\begin{figure}[tb]
\includegraphics*[width=12.0cm]{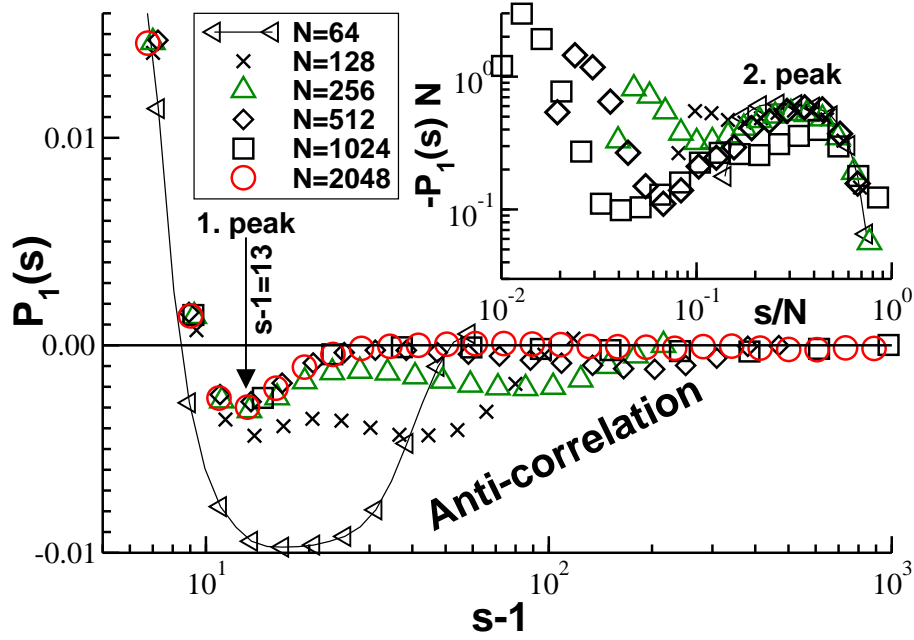}
\caption{Bond-bond correlation function $P_1(s) = \langle \vec{e}_n \cdot \vec{e}_m \rangle$  
{\em vs.} curvilinear distance $s-1$.
Main panel: The first Legendre polynomial shows an anti-correlation at $s-1 \approx 13$
due to the local wrapping of the chain.
Inset: A second anti-correlation peak is visible at $s \approx N/2$ which is caused by
the reflection of the confined chain. Due to Eq.~(\ref{eq_P1Rs}) this peak is related to 
the decay of the reduced subchain size $\Rend^2(s)/s^{1/2}$ visible in Fig.~\ref{fig_RsB}.
\label{fig_P1} 
}
\end{figure}

\begin{figure}[tb]
\includegraphics*[width=12.0cm]{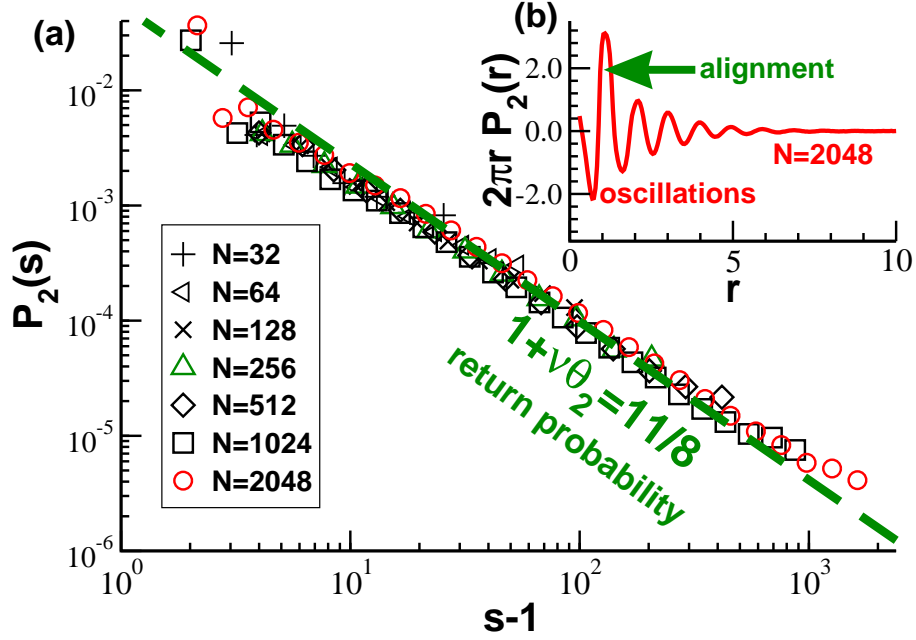}
\caption{Second Legendre polynomial 
$P_2 = \langle \left(\vec{e}_n \cdot \vec{e}_m\right)^2 \rangle -1/2$ 
as a function of curvilinear distance $s-1$ {\bf (a)}
and as a function of intrachain distance $r$ {\bf (b)}. 
{\bf (a)} $P_2(s)$ decays over two orders in magnitude as a power law (dashed line)
with an exponent $1+\nu\theta_2=11/8$ in agreement with a return probability given by Eq.~(\ref{eq_preturn}). 
{\bf (b)} $2\pi r P_2(r)$ oscillates strongly with an exponentially decreasing amplitude.
Being dominated by a nematic peak at the origin ($r \approx 1$) it acts as a $\delta$-function.
\label{fig_P2} 
}
\end{figure}

\begin{figure}[tb]
\includegraphics*[width=12.0cm]{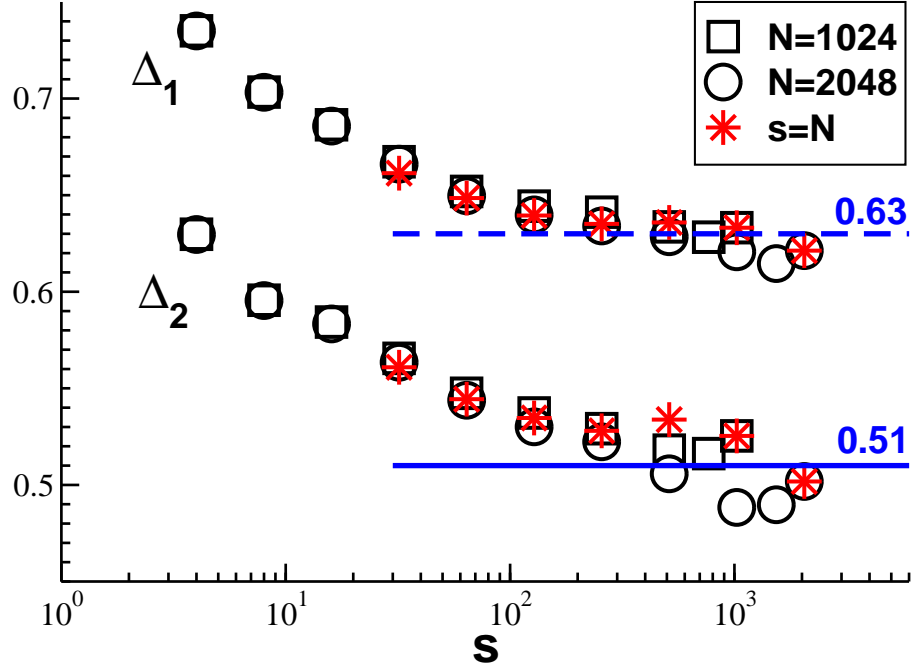}
\caption{Aspherity of chains (stars) and subchains for $N=1024$ (squares) and $N=2048$ (spheres)
using log-linear coordinates. The aspherity is characterized by the moments $\Delta_1$ and $\Delta_2$ 
of the eigenvalues $\lambda_1$ and $\lambda_2$ of the inertia tensor discussed in the main text. 
Both moments decay weakly with $s$ and $N$. The horizontal lines are guides to the eye indicating
possible plateau values for asymptotically long chains.
\label{fig_asph}
}
\end{figure}

As shown in Fig.~\ref{fig_snapshot}, 2D chains adopt irregular shapes with perimeters 
{\em not} appearing to be smooth, i.e. characterized by a line dimension $\dc=1$, 
but clearly fractal ($2 > \dc > 1$) \cite{Mandelbrot}.
In this subsection we analyze quantitatively this visual impression confirming the announced
key result Eq.~(\ref{eq_keyLs}) for the average perimeter length $L(s)$ of chains ($s=N$) and 
subchains ($s \le N$).  

\begin{figure}[tb]
\includegraphics*[width=12.0cm]{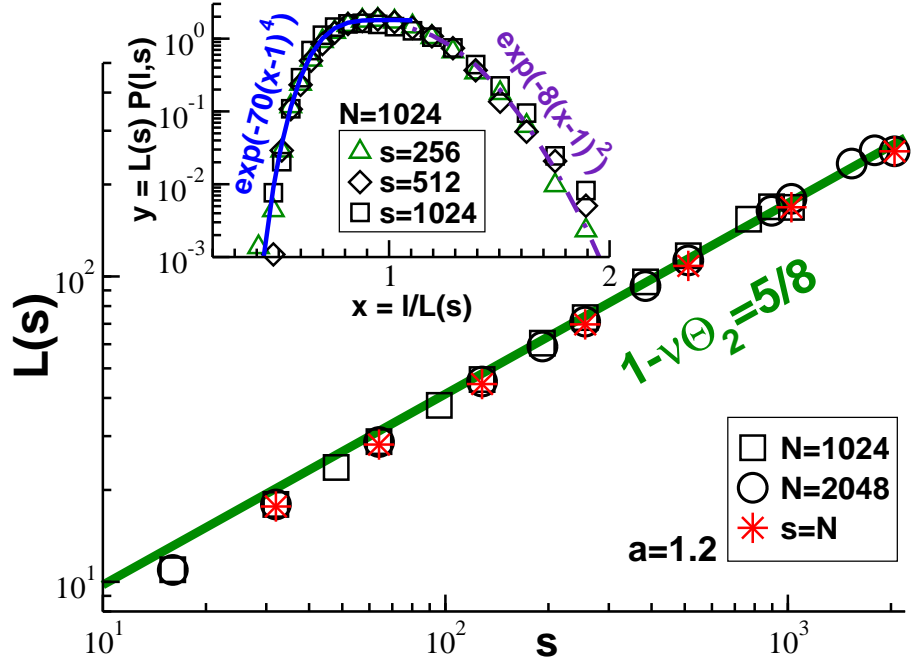}
\caption{Mean perimeter length $L(s) = \langle l \rangle$ (main panel) 
and perimeter length distribution $P(l,s)$ (inset) for $a=1.2$.
In agreement with Eq.~(\ref{eq_keyLs}) the perimeter scales as 
$L(s) \sim s^{1-\nu\Theta_2} \sim s^{5/8}$ (bold line). 
Stars refer to perimeters of chains ($s=N$),
open symbols to subchains of lengths $s \le N$ in chains of length $N=1024$ (squares) and $N=2048$ (spheres).
The histograms for different $s$ collapse if rescaled using $L(s)$ as presented in the inset for 
different segment sizes and $N=1024$.
The distributions are lopsided decaying much sharper for small-perimeter segments ($l/L(s) \ll 1$).
The exponential cut-offs indicated for both limits are phenomenological fits.
\label{fig_Ls}
}
\end{figure}

%
We define a perimeter monomer as having at least one monomer not belonging to the same chain or 
subchain closer than a reference distance $a \approx 1$ essentially set by the
monomer density (see below). Specifically, we have used $a=1.2$ in 
Figs.~\ref{fig_snapshot}, \ref{fig_snapblob} and \ref{fig_Ls} and 
for the data listed in the Table. 
The number of such perimeter monomers is called $l(s)$, its mean number $L(s) = \la l(s) \ra$
with $L(s)/s \le 1$ being the probability that a monomer of a subchain is on its perimeter.
(Note that for a continuous chain model a slightly different ``box counting" method must be
used to obtain a finite perimeter length \cite{Mandelbrot}.)
The main panel of Fig.~\ref{fig_Ls} presents $L(s)$ using the same symbols as in Fig.~\ref{fig_Rs}.
All data collapses on the same master curve, confirming nicely the announced exponent 
$1-\nu\Theta_2=5/8$ (dashed line) and thus a fractal line dimension $\dc = (5/8) /\nu = 5/4$.
This result holds provided that chain and subchain lengths are not too small ($N, s \gg 50$). 
%

%
%
%
Having just confirmed Eq.~(\ref{eq_keyLs}) numerically we have still to give a theoretical 
argument to show where it stems from. 
Using the return probability $p_r(s) \sim 1/s^{1+\nu\Theta_2}$ measured in Fig.~\ref{fig_P2}
a simple scaling argument can be given following Semenov and Johner \cite{ANS03}.
The key point is that a monomer in a long subchain cannot ``distinguish'' if the contact is realized
through the backfolding of its own subchain or by another subchain of length $s$.
Since the probability of such a contact, $L(s)/s$, must be proportional to $p_r(s)$
times the number $s$ of monomers in the second subchain we have 
\begin{equation}
L(s)/s \sim p_r(s) \times s \sim s^{-\nu\Theta_2} \mbox{ for } 1 \ll s \leq N
\label{eq:pint}
\end{equation}
which is identical to Eq.~(\ref{eq_keyLs}).
An alternative, but related derivation will be given in Section~\ref{sub_Fq}.

%
The fluctuations of the perimeter length are characterized in the inset of Fig.~\ref{fig_Ls}.
We present here the histograms $P(l,s)$ of the number of perimeter monomers for different subchains of lengths 
$s$ for chains of length $N=1024$. Assuming that the first moment $L(s)$ of the histogram sets the only scale
all histograms are successfully brought to a scaling collapse.
Please note that the fractality of the perimeter does not imply that the histograms have to be broad.
In fact, they decay rather rapidly as indicated by the two phenomenological fits and all moments
of the distributions exist. The configurations corresponding to small perimeters are rather
strongly suppressed (solid line). At variance to this, the decay for $l(s) - L(s) \gg 0$ is found to
be successfully described by a Gaussian (dash-dotted line). The perimeter fluctuations of different 
contour sections of these configurations are apparently only weakly coupled, if at all.
%

\begin{figure}[tb]
\includegraphics*[width=12.0cm]{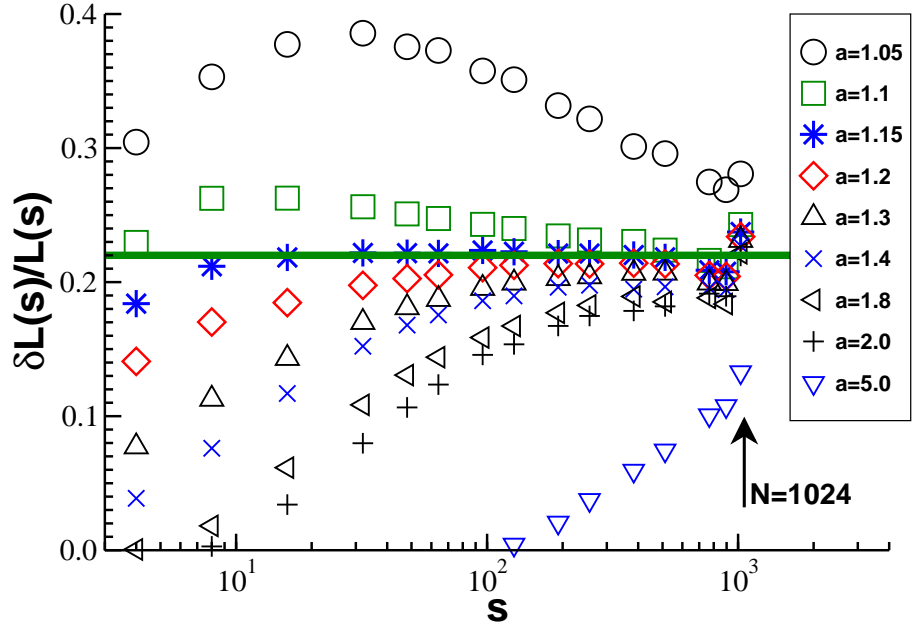}
\caption{Relative error $\delta L(s)/L(s)$ as a function of $s$ for chains of length $N=1024$
and for different distances $a$ as indicated. 
The relative error increases for small $s$ and too large $a$-values.
It decreases for $a < 1.15$ since these values are too small for the given density
and not all perimeter monomers are detected. The relative error becomes constant
for large $s$ irrespective of the value of $a$.
Due to chain end effects the relative error increases slightly for $s \approx N$.
\label{fig_dLL}
}
\end{figure}

%
The influence of the distance $a$ used to define a perimeter monomer is investigated 
in Fig.~\ref{fig_dLL} where we present the ``relative error" 
$\delta L(s)/L(s) \equiv \sqrt{\la l^2\ra / \la l \ra^2 - 1}$ 
of the distributions $P(l,s)$. (The relative error for $s=N$ is listed in the Table.)
For clarity, only data for one chain length $N=1024$ is given for several $a$ as indicated in the figure.
Obviously, if $a$ is too large ($a \gg R(s)$) all subchain monomers are considered to be perimeter monomers,
$l \approx s$, and the perimeter length cannot fluctuate ($\delta L(s)/L(s) \approx 0$). With increasing $s$
the fluctuations increase first ($a > 1.15$) and level then off in the limit of large $s$, i.e. 
$\delta L(s)/L(s) \sim s^0$ in agreement with the scaling found in the inset of Fig.~\ref{fig_Ls}. 
If, on the other hand, $a$ is too small ($a < 1.15$) not all monomers clearly on the contour
are detected, as can be checked by looking at snapshots similar to Fig.~\ref{fig_snapshot}.
In this case the fluctuations first decay until the subchain length is sufficiently large
that the number of detected perimeter monomers becomes proportional to the true number.
Hence, the relative errors of too small and too large $a$ essentially merge for large $s$ or become parallel. 
The specific value of $a$ is thus inessential from the scaling point of view.
However, computationally it is important to choose a parameter $a$ 
allowing to probe the asymptotic scaling behavior for as broad an $s$-range as possible.
It is for this technical reason that $a=1.2$ has been chosen above.

\begin{figure}[tb]
\includegraphics*[width=12.0cm]{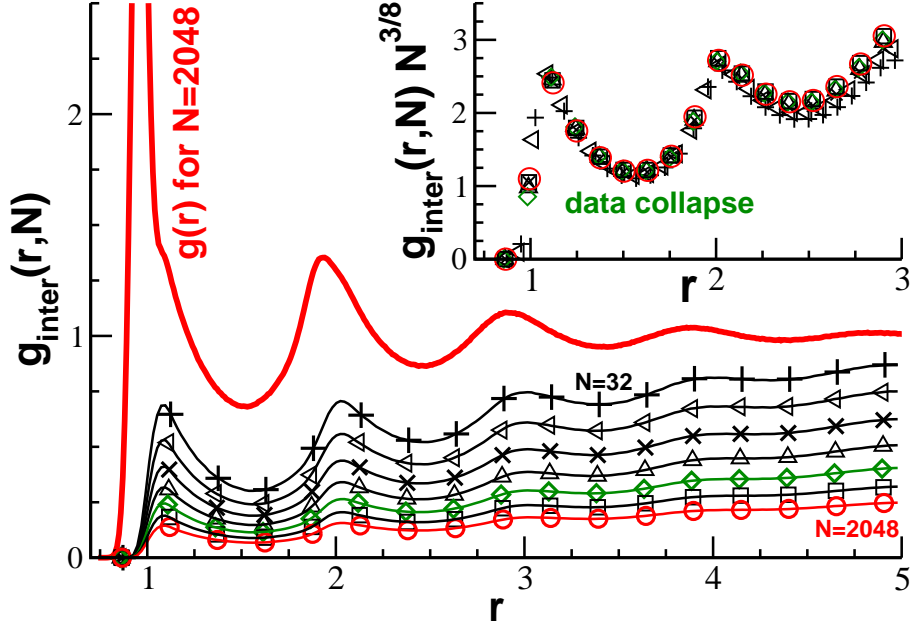}
\caption{Radial pair correlation function $\ginter(r,N)$ between monomers on different chains
using the same symbols as in Fig.~\ref{fig_P2} for the chain length $N$. 
Also indicated is the pair correlation function $g(r)$ between all monomers for $N=2048$ (solid line).
Measuring the probability that a monomer is close to the chain perimeter $\ginter(r,N)$ decreases strongly with $N$.
This decay is perfectly described by $\ginter(r,N) \sim N^{-\nu \Theta_2} = 1/N^{3/8}$ 
as shown by the scaling collapse of the data presented in the inset.
\label{fig_grinter}
}
\end{figure}

%
A method allowing to verify the fractal dimension of the chain contour
not requiring such an artificial parameter is presented in Fig.~\ref{fig_grinter}.
We show here the radial pair correlation function $\ginter(r,N)$ between monomers
on {\em different} chains as a function of the distance $r$ between the monomers.
The bold line indicates the pair correlation function $g(r)$ between {\em all} monomers
already presented in Fig.~\ref{fig_gr}. The same normalization is used for $\ginter(r,N)$ 
as for $g(r)$, i.e. $\ginter(r,N) \to 1$ for large distances where both monomers
must necessarily stem from different chains. Obviously, $g(r) \ge \ginter(r,N)$ for
all distances $r$. For small distances $r \approx 1$ the pair correlation function $\ginter(r,N)$ 
measures the probability that two monomers from different chains are in contact.
We remind that for {\em open} chains, e.g. self-avoiding chains in 3D melts, $\ginter(r,N)$
becomes rapidly chain length independent. Our chains are {\em compact}, however, and only
a fraction of the chain monomers is close to its contour. Hence, $\ginter(r,N)$ 
must decrease with $N$ for $r \approx 1$. This is clearly confirmed by the data presented in the main panel.
From Eq.~(\ref{eq:pint}) we know already the probability for two monomers from different chains
to be close to each other, i.e. for both monomers to be close to the chain contour.
One expects thus to find
\begin{equation}
\ginter(r,N) \sim L(N)/N  \sim N^{-\nu \Theta_2} \sim N^{-3/8}
\label{eq_grinter}
\end{equation}
for small distances $r$ of the order of a few monomer diameters. This scaling is perfectly demonstrated 
by the data collapse presented for all chain lengths in the inset of Fig.~\ref{fig_grinter}.
%

\subsection{Intrachain structure factor}
\label{sub_Fq}
\begin{figure}[tb]
\includegraphics*[width=12.0cm]{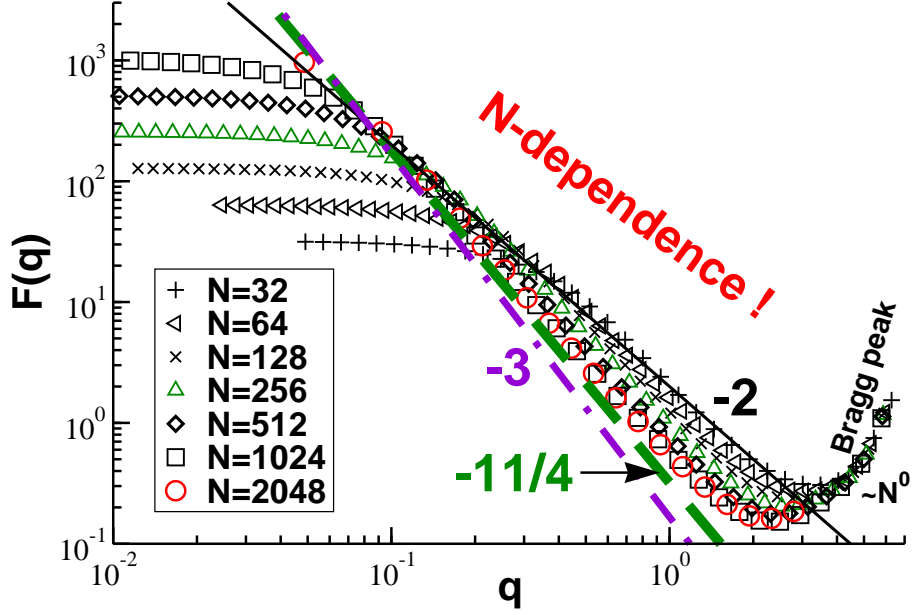}
\caption{Intramolecular structure factor $F(q)$ as a function of wave vector $q$
for different chain lengths $N$ as indicated. In striking contrast to 3D melts
$F(q)$ becomes chain length independent only for wave vectors corresponding to the monomer scale (``Bragg peak"). 
To characterize the decay in the intermediate wave vector regime our data is compared with three
power-law exponents $-2$, $-11/4$ and $-3$ indicated by the thin, the bold dashed line and the
dash-dotted line, respectively. 
The first exponent corresponds to Eq.~(\ref{eq_Fqopen}),
the second exponent to Eq.~(\ref{eq_keyFq})
and the last exponent to Eq.~(\ref{eq_Fqsurface}), 
i.e. to the Porod scattering of a compact 2D object with smooth surface ($d=2,\dc=1$). 
\label{fig_Fq}
}
\end{figure}

%
%
Neither the intrachain size distributions $G_i(r,s)$, nor the 
bond-bond correlation functions $P_1(s)$ and $P_2(s)$ 
or the contour length $L(s)$ are readily accessible experimentally,
at least not for classical small-monomer polymers which cannot be 
visualized directly by means of fluorescence microscopy or atomic 
force microscopy.
It is thus important to demonstrate that $\Theta_2$ is measurable 
in principle from an analysis of the intrachain structure factor 
$F(q) = \frac{1}{N} \sum_{n,m=1}^{N} \langle \exp\left[i \vec{q} \cdot (\vec{r}_n-\vec{r}_m) \right]\rangle$
as announced in Eq.~(\ref{eq_keyFq}) in the Introduction and 
as shown now in Fig.~\ref{fig_Fq} and Fig.~\ref{fig_Fqkratky}.
%

%
Figure~\ref{fig_Fq} presents the unscaled structure factors $F(q)$ as a function of the wave vector $q$
for a broad range of chain lengths $N$ as indicated. As one expects, $F(q)$ becomes constant for
very small wave vectors ($F(q) \to N$), decreases in an intermediate wave vector regime 
($\Rgyr(N) \ll 1/q \ll 1$) and shows finally the non-universal monomer structure 
for large $q$ comparable to the inverse monomeric size (``Bragg peak"). 
The first striking result of this plot is that $F(q)$ does {\em not} become chain length independent 
in the intermediate wave vector regime as it does for (uncollapsed) polymer chains in 3D.
The second observation to be made is that with increasing chain length the decay becomes {\em stronger} 
than the power-law exponent $-2$ indicated by the thin line corresponding to Gaussian chain statistics.   

%
%
Since for an {\em open} polymer-like aggregate or cluster of inverse fractal dimension $\nu$
without any sharp surface the structure factor must indeed scale as \cite{DegennesBook,BenoitBook} 
\begin{equation}
F(q) \sim N^0 q^{-1/\nu} \mbox{ for } q \gg 1/\Rgyr(N)
\label{eq_Fqopen}
\end{equation}
several authors \cite{carmesin90,mai99,pakula02,yethiraj03} have argued 
that an exponent $-2$ should be observed for 2D polymer melts. However, Eq.~(\ref{eq_Fqopen}) 
does not hold for {\em compact} structures where strong composition fluctuations 
(of the labeled monomers of the reference chain with respect to unlabeled monomers) 
at a thus well-defined surface or perimeter must dominate the structure factor
leading to a ``generalized Porod scattering" \cite{BenoitBook}.  
Since the exponent $\nu=1/2$ for 2D melts does not refer to their Gaussian open chain statistics
but rather to their compactness, Eq.~(\ref{eq_Fqopen}) is thus inappropriate.
%
Quite generally, the scattering intensity $N F(q)$ of compact objects is known to be proportional to their 
``surface" $L(N)$ which implies \cite{BenoitBook,bale84,bray88}
\begin{equation}
N F(q) \approx N^{2} / \left(q R(N) \right)^{2d-\dc}
\sim R(N)^{\dc} \sim L(N)
\label{eq_Fqsurface}
\end{equation}
where we have used that $R(N) \sim N^{1/d}$ and $R(N)^{\dc} \sim L(N)$.
For a 2D object with a smooth perimeter, i.e. for $\dc=1$, this corresponds to the 
well-known Porod scattering $F(q) \sim q^{-3}$. As indicated by the dash-dotted line in Fig.~\ref{fig_Fq},
this yields a too strong decay not compatible either with our data. Obviously, this is to be expected 
since we already know from Sec.~\ref{sub_Ls} that the perimeter is fractal ($\dc > 1$)
and the power-law slope $-3$ must be a lower bound to our data.
If we assume, on the contrary, in Eq.~(\ref{eq_Fqsurface}) a fractal line dimension 
$\dc=d-\Theta_2$, as demonstrated analytically in Eq.~(\ref{eq:pint}), 
this yields directly  the key result Eq.~(\ref{eq_keyFq}) anticipated in the Introduction.
Using $\Theta=3/4$ we thus have $F(q) \sim q^{-11/4}$ as indicated by the dashed line.
This power law gives a reasonable fit for the largest chains we have computed.

%
The representation of the structure factor used in Fig.~\ref{fig_Fq} is not the best one to 
check the asymptotic power-law exponents and does not allow to verify the $N$-scaling implied 
by Eq.~(\ref{eq_Fqsurface}). We have thus replotted our data in Fig.~\ref{fig_Fqkratky} 
using a Kratky representation with vertical axis $y=(F(q)/N) Q^2$ and a reduced wave vector 
$Q = q \Rgyr(N)$. Using the measured radius of gyration $\Rgyr(N)$ given in the Table 
this obviously allows to collapse all data in the Guinier regime for $Q \ll 1$
where \cite{DoiEdwardsBook}
\begin{equation}
y(Q) = Q^2 \left(1 - \frac{Q^2}{d}\right).
\label{eq_Guinier}
\end{equation}
The observed data collapse is, however, much broader in $Q$ and the more the larger the chain length.
The deviations observed for large $Q$ are due to (chain length independent) physics 
on scales corresponding to the monomer size (``Bragg peak") already seen in Fig.~\ref{fig_Fq}.
The Debye formula for Gaussian chains \cite{DoiEdwardsBook} is given by the thin line which 
becomes constant in this representation for $Q \gg 1$ in agreement with Eq.~(\ref{eq_Fqopen}). 
At variance to this, a striking decay of $y(Q)$ is observed over a decade in $Q$
confirming observations by Yethiraj \cite{yethiraj03} and Cavallo {\em et al.}
\cite{CMWJB05} using much shorter chains.
Due to the scaling of $y(Q)$ as a function of $Q$ this decay implies the chain length 
dependence seen for the unscaled structure factor in Fig.~\ref{fig_Fq}.
With increasing chain length our data approaches systematically the power law $y(Q) \sim Q^{-\Theta_2}$
given by Eq.~(\ref{eq_keyFq}) and indicated by the dashed line. Even longer chains obviously are 
warranted to unambiguously show the predicted asymptotic exponent $-3/4$ in a computer experiment.

%
In the preceding two paragraphs we have used the fractal line dimension $\dc=d-\Theta_2$
derived {\em via} Eq.~(\ref{eq:pint}) together with the generalized Porod scattering scaling 
Eq.~(\ref{eq_Fqsurface}) to demonstrate the key result Eq.~(\ref{eq_keyFq}).
Interestingly, the structure factor can be computed directly {\em without} the scaling 
argument Eq.~(\ref{eq:pint}) using that $\Ge(r,s) \approx G_2(r,s)$ for $s \ll N$ as discussed
at the end of Sec.~\ref{sub_theta}. 
For asymptotically long chains the structure factor thus can be well approximated as
\begin{equation}
F(q) \approx \frac{1}{N} \int_0^N ds \ 2(N-s) G_2(q,s) 
\label{eq_Gs2Fq}
\end{equation}
using the Fourier transform $G_2(q,s)$ of $G_2(r,s)$. Within the Redner-des Cloizeaux approximation,
Eq.~(\ref{eq_Redner}), this yields an analytic formula, Eq.~(\ref{ap_Fq1}), given in the Appendix. 
Readily computed numerically, this theoretical prediction is represented by the solid line in Fig.~\ref{fig_Fqkratky}.
Since Eq.~(\ref{eq_Redner}) and Eq.~(\ref{eq_Gs2Fq}) are both approximations this result is not strictly rigorous.
However, by construction our formula must yield the Guinier regime, Eq.~(\ref{eq_Guinier}), for small $Q=q\Rgyr(N)$
and since for large $Q$ only the $\Theta_2$-exponent matters for large $N$ it is only around the
hump $Q \approx 2$ where deviations could be relevant. Fig.~\ref{fig_Fqkratky} shows that in practice
our approximation agrees well for all $Q$ as long as the wave vector $q$ does not probe local physics
(Bragg regime).

As shown in the Appendix, the Redner-des Cloizeaux approximation Eq.~(\ref{ap_Fq1}) reduces to a 
power law for wave vectors $Q \gg 1$ corresponding to the power-law regime of Eq.~(\ref{eq_theta}),
\begin{equation}
y(Q) \approx
\frac{2}{\Gamma(2-\Theta_2/2)} \left(\frac{3}{2+\Theta_2}\right)^{-(1+\Theta_2/2)} Q^{-\Theta_2}
\approx \frac{1.98}{Q^{3/4}}, 
\label{eq_Fqpower}
\end{equation}
in agreement with the key Eq.~(\ref{eq_keyFq}) given in the Introduction.
Equation~(\ref{eq_Fqpower}) is represented by the dashed lines in Fig.~(\ref{fig_Fq}) 
and Fig.~(\ref{fig_Fqkratky}). 
%
Comparing Eq.~(\ref{eq_Fqsurface}) with Eq.~(\ref{eq_Fqpower}) demonstrates that 2D melts are characterized
by a fractal line dimension 
\begin{equation}
\dc = d - \Theta_2 = 5/4.
\label{eq_dcalter}
\end{equation}
Hence, using a slightly more physical route as the scaling argument given in Sec.~\ref{sub_Ls}
we have confirmed the fractal line dimension of the chain perimeter $L(N)$.
By labeling only the monomers of sub-chains 
(which corresponds to a scattering amplitude $s F(q) \sim L(s) \sim R(s)^{\dc}$)
the above argument is readily generalized to the perimeter 
length $L(s)$ of arbitrary segments of length $s \le N$.

\section{Conclusion}
\label{sec_concl}

%
Using scaling arguments and molecular dynamics simulation of a well-known model Hamiltonian
we investigated various static  properties of linear polymer melts in two dimensions.
We have shown that the chains adopt compact conformations ($\nu=1/d=1/2$).
Due to the segregation of the chains the ratio of end-to-end distance $\Rend(N)$ and gyration 
radius $\Rgyr(N)$ becomes $\Rend^2(N)/\Rgyr^2(N) \approx 5.3 < 6$ (Fig.~\ref{fig_RsB})
and the chains are more spherical than Gaussian phantom chains (Fig.~\ref{fig_asph}).
More importantly, it is shown that the irregular chain contours can be characterized by a 
fractal line dimension $\dc=d-\Theta_2=5/4$ (Figs.~\ref{fig_Ls} and \ref{fig_grinter}). 
This key result has been demonstrated analytically using two different scaling arguments 
given in Sec.~\ref{sub_Ls} and Sec.~\ref{sub_Fq}, both based on the numerically tested 
power-law scaling of the intrachain size distribution $\Ge(r,s) \approx G_2(r,s) \sim r^{\Theta_2}$ 
for small distances $r \ll \Rend(s)$ with $\Theta_2=3/4$ (Fig.~\ref{fig_theta}).
Compactness and perimeter fractality repeat for subchains of arc-lengths $s$ down to 
a few monomers due to the self-similar structure of the chains 
(Figs.~\ref{fig_snapblob},\ref{fig_Rs} and \ref{fig_Ls}).
Measuring directly the return probability of the chain after $s$ steps, 
the second Legendre polynomial $P_2(s)$ of the bond vectors decays
as $P_2(s) \sim 1/s^{1+\nu\Theta_2}$ (Fig.~\ref{fig_P2}). 
Interestingly, as implied by the generalized Porod scattering of a compact object with 
fractal ``surface", Eq.~(\ref{eq_Fqsurface}), the predicted fractal line dimension 
should in principle be accessible experimentally from the power-law scaling, Eq.~(\ref{eq_keyFq}), 
of the intrachain structure factor $F(q)$ in the intermediate wave vector regime. 
Computationally very demanding systems with chain lengths up to $N=2048$ have been required 
to test the proposed scaling of the structure factor (Fig.~\ref{fig_Fqkratky}).
%

%
We would like to stress that our results are not restricted to a particular melt density,
but should also hold for all densities provided that the chains are sufficiently long
to allow a renormalization of all length scales in terms of semidilute blobs \cite{DegennesBook}.
This is of some interest since chain conformations of semidilute 2D solutions of large-monomer
polymers (such as DNA or brushlike polymers) are experimentally better accessible than dense melts 
\cite{mai99,dobrynin07}. 
Obviously, these macromolecules are rather rigid and in view of the typical molar masses currently used, 
deviations are to be expected from the asymptotic chain length behavior we focused on.
Following previous computational work
\cite{carmesin90,rutledge97,pakula02,yethiraj03,yethiraj05,deutsch05}
it should thus be rewarding to reinvestigate the scaling of flexible and {\em semiflexible} chains in 
2D semidilute solutions to see how finite-$N$ effects may systematically be taken into account.

Interestingly, the fractality of the perimeter precludes a finite line tension and the shape 
fluctuations of the segments are not suppressed exponentially \cite{carmesin90}, but may occur 
by advancing and retracting ``lobes'' in an ``amoeba-like'' fashion.
This opens the possibility for a relaxation mechanism, specific to 2D polymer melts, 
in which energy is dissipated by friction at the boundary between subchains. 
Following a suggestion made recently \cite{ANS03},
the longest relaxation time $\tau(s)$ of a chain segment should, hence, scale as
\begin{equation} 
\tau(s) \sim L(s)^3 \sim s^{\alpha}
\mbox{ with } \alpha = 3 (1-\nu \Theta_2)=15/8
\label{eq_taus}
\end{equation}
rather than with $\alpha=2$
as predicted by the Rouse model which is based on Gaussian chain statistics \cite{DoiEdwardsBook}.
As Gaussian chain statistics is inappropriate to describe conformational properties of
2D melts, there is no reason why a modeling approach based on this statistics may allow to describe,
e.g., the composition fluctuations at the chain contour. Since the latter can in principle be probed
experimentally using the dynamical intrachain structure factor $F(q,t)$ \cite{BenoitBook} 
this is an important issue we are currently investigating. 

\begin{acknowledgments}
We thank the ULP, the IUF, the Deutsche Forschungsgemeinschaft (Grant No. KR 2854/1-1), 
and the ESF-STIPOMAT programme for financial support. A generous grant of computer time 
by GENCI-IDRIS (Grant No. i2009091467) is also gratefully acknowledged.
We are indebted to A.N. Semenov (Strasbourg), S.P. Obukhov (Gainesville), 
M. M\"uller (G\"ottingen), M. Aichele (Frankfurt) and A. Cavallo (Salerno) 
for helpful discussions.
\end{acknowledgments}

\appendix
\section{Calculation of intrachain structure factor}
\label{app_Fq}

The intramolecular structure factor $F(q)$ may be rewritten generally as \cite{WBM07}
\begin{equation}
F(q) 
= \frac{1}{N} \int_0^N ds \ 2(N-s) \Ge(q,s)
\label{ap_FqGe}
\end{equation}
using the Fourier transform $\Ge(q,s)$ of the two-point intramolecular correlation function 
$\Ge(r,s)$ averaging over all pairs of monomers $(n,m=n+s-1)$ discussed in Sec.~\ref{sub_theta}.
As we have seen in Sec.~\ref{sub_theta}, $\Ge(r,s)$ is well approximated by the distribution 
$G_2(r,s)$ for $s \ll N$. For asymptotically long chains it is justified to 
neglect chain-end effects ($s \to N$), i.e. physics described by the contact exponents $\Theta_0$
and $\Theta_1$. Assuming thus translational invariance along the chain contour the structure 
factor is given approximately by 
\begin{equation}
F(q) \approx \frac{1}{N} \int_0^N ds \ 2(N-s) G_2(q,s),
\label{ap_FqG2}
\end{equation}
the factor $2 (N-s)$ counting the number of equivalent monomer pairs separated by an arc-length $s$. 
Using the Redner-des Cloizeaux approximation, Eq.~(\ref{eq_Redner}), for $i=2$ we compute first the 
2D Fourier transform
\begin{equation}
G_2(q,s) 
         = \int_0^{\infty} c_2 x^{\Theta_2} e^{-k_2 x^2} \ 2 \pi x dx J_0(q x) \label{ap_G2Redner} 
\end{equation}
with $2\pi J_0(z) = 2 \int_0^{\infty} \cos(z \cos(\theta)) d\theta$ 
being an integer Bessel function \cite{abramowitz} and 
$x=r/R_2(s)= r/\be s^{1/2}$, $\Theta_2=3/4$, $k_2=1+\Theta_2/2$, $c_2=k_2^{k_2}/\pi \Gamma(k_2)$
as already defined in Sec.~\ref{sub_theta}.
As can be seen from Eq.~(11.4.28) of Ref.~\cite{abramowitz}, this integral is given by
a standard confluent hypergeometric function, the Kummer function $M(a,b,-z)$,
\begin{equation}
G_2(q,s) = M(1+\Theta_2/2,1,-z)
\label{ap_G2M}
\end{equation}
with $z = q^2\be^2 s/4 k_2$. 
According to Eq.~(13.1.2) and Eq.~(13.1.5) of \cite{abramowitz} the Kummer function can be expanded as
\begin{eqnarray}
M(a,b,-z) & \approx & 1 - \frac{a z}{b} \mbox{ for } |z| \ll 1, \label{ap_Msmallz} \\
M(a,b,-z) & \approx & \frac{\Gamma(b)}{\Gamma(b-a)} z^{-a} \mbox{ for } z \gg 1.
\label{ap_Mlargez}
\end{eqnarray}
Using Eq.~(\ref{ap_G2M}) this yields, respectively, the small and the large wave vector 
asymptotic behavior of the Fourier transform of $G_2(r,s)$
\begin{eqnarray}
G_2(q,s) & \approx & 1 - (1+\Theta_2/2) z \mbox{ for } z \ll 1, \label{ap_G2smallz} \\
G_2(q,s) & \approx & \frac{\Gamma(1)}{\Gamma(-\Theta_2/2)} z^{-(1+\Theta_2/2)} 
\sim q^{-(2+\Theta_2)} \mbox{ for } z \gg 1. \label{ap_G2largez}
\end{eqnarray}
Note that Eq.~(\ref{ap_G2smallz}) implies $G_2(q=0,s) = 1$ as one expects due to the normalization of $G_2(r,s)$.

\begin{figure}[tb]
\includegraphics*[width=12.0cm]{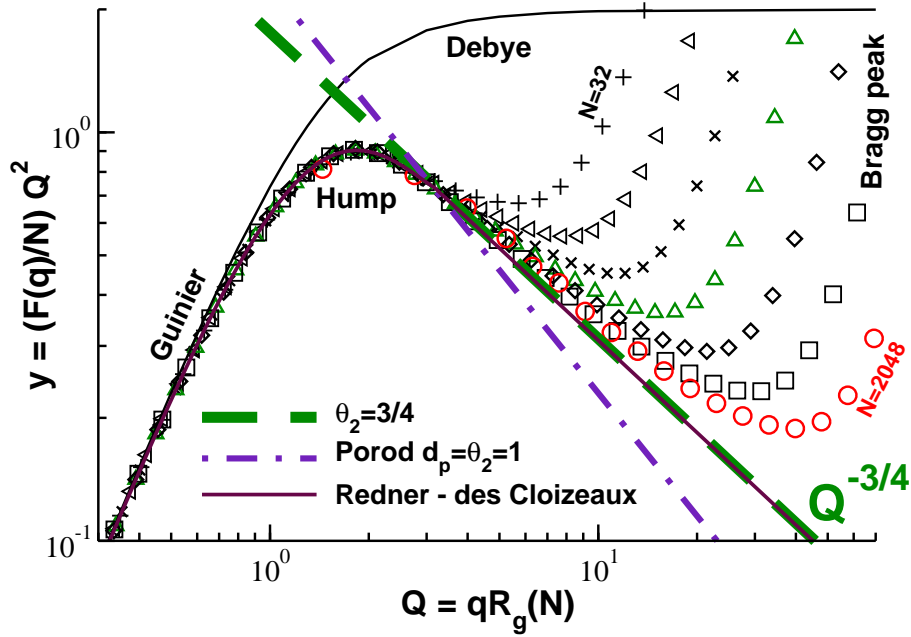}
\caption{
Kratky representation of the structure factor $F(q)$
tracing $y=(F(q)/N) Q^2$ as a function of the reduced
wave vector $Q=q\Rgyr(N)$
for different $N$ using the same symbols as in Fig.~\ref{fig_Fq}. 
The Debye formula (top thin line) corresponds to a plateau for $Q \gg 1$. 
At variance to this a strong non-monotonous behavior is revealed by our data
which approaches with increasing $N$ a power law exponent $-\Theta_2=-3/4$ 
(dashed line) corresponding to a compact object of fractal line dimension $\dc =d - \theta_2 = 5/4$.
The Porod scattering for a compact 2D object with smooth perimeter is given by the
dash-dotted line, the Fourier transform of the Redner-des Cloizeaux approximation by 
the solid line. 
%
\label{fig_Fqkratky}
}
\end{figure}

After integrating over $s$ following Eq.~(\ref{ap_FqG2}) and defining $Z = q^2\be^2 N/4 k_2$ 
one obtains for the Guinier regime of the structure factor 
\begin{equation}
F(q) \approx N \left( 1 - \frac{1+\Theta_2/2}{3} Z \right) \mbox{ for } Z \ll 1, 
\label{ap_Fsmallq} 
\end{equation}
i.e. according to Eq.~(\ref{eq_Guinier}) we have, as one expects, 
\begin{equation}
\Rgyr^2(N) = \frac{1}{6} \be^2 N \frac{1+ \Theta_2/2}{k_2} = \frac{\be^2 N}{6}.
\label{ap_Rg}
\end{equation}
This is consistent with Eq.~(\ref{eq_Rend2Rgyr}) and the ratio $(\be/\bg)^2=6$ 
with $\be$ determined from subchains with $s \ll N$, Eq.~(\ref{eq_beRs}).
Eq.~(\ref{ap_Rg}) is of course slightly at variance with the measured ratio 
$(\Rend(N)/\Rgyr(N))^2=(\bestar/\bg)^2 < 6$ 
due the end-effects not taken into account in Eq.~(\ref{ap_FqG2}).
The power law behavior of the structure factor for large wave vectors
announced in Eq.~(\ref{eq_keyFq}) is obtained by
integrating Eq.~(\ref{ap_G2largez}) with respect to $s$. This gives
\begin{equation}
F(q) \approx \frac{2 N}{\Gamma(2-\Theta_2/2)} Z^{-(1+\Theta_2/2)} 
\sim N^{-\Theta_2/2} q^{-(2+\Theta_2)} 
\mbox{ for } Z \gg  1.
\label{ap_Flargeq}
\end{equation}
Obviously, it is also possible to directly integrate Eq.~(\ref{ap_G2M}) with respect to $s$ 
according to Eq.~(\ref{ap_FqG2}). This yields the complete Redner-des Cloizeaux approximation
of the structure factor
\begin{eqnarray}
\frac{F(q)}{N}  & \approx & 
2 M\left(1+\frac{\Theta_2}{2},2,-Z\right) \nonumber \\
& - & M\left(1+\frac{\Theta_2}{2},3,-Z\right) \nonumber \\
& + &  
\frac{1}{3} \left(1+\frac{\Theta_2}{2}\right) Z \ 
M\left(2+\frac{\Theta_2}{2},4,-Z\right)
\label{ap_Fq1}
\end{eqnarray}
which can be computed numerically.
Using again the expansions of the Kummer function, Eq.~(\ref{ap_Msmallz}) and Eq.~(\ref{ap_Mlargez}), 
one verifies readily that Eq.~(\ref{ap_Fq1}) yields the asymptotics for small and large wave vectors
already given above. 
It is convenient from the scaling point of view to replace the variable $Z$ used above
by the reduced wave vector $Q = q \Rgyr(N)$ substituting
\begin{equation}
Z \Longrightarrow \frac{6}{4} Q^2 (1+\Theta_2/2) = \frac{12}{11} Q^2,
\label{ap_Zreplace}
\end{equation}
as suggested by Eq.~(\ref{ap_Rg}). This gives the curve represented by the 
solid line in Fig.~\ref{fig_Fqkratky}. Eq.~(\ref{ap_Flargeq}) reexpressed
in these terms is given by Eq.~(\ref{eq_Fqpower}) in the main text.
Note that due to this substitution the Guinier limit of the Redner-des Cloizeaux approximation 
of the structure factor is correct by construction.


%
\newpage

\begin{table}[t]
\begin{tabular}{|c||c|c|c|c|c|c|c|c|c|}
\hline
$N$   & $\Rend$ & $\Rgyr$ &$\left(\frac{\Rend}{\Rgyr}\right)^2$ & $\frac{\la \lambda_1\ra}{\la \lambda_2\ra}$ 
& $\Delta_1$ & $\Delta_2$ & $\frac{L(N)}{N}$ & $\frac{\delta L(N)}{L(N)}$ \\ \hline
32    & 8.1     & 3.4     & 5.7 & 4.9 & 0.66 & 0.56      & 0.55    & 0.21 \\ 
64    & 11.7    & 5.0     & 5.4 & 4.7 & 0.65 & 0.54      & 0.44    & 0.22 \\ 
128   & 16.7    & 7.2     & 5.4 & 4.6 & 0.64 & 0.54      & 0.35    & 0.23 \\ 
256   & 23.8    & 10.3    & 5.3 & 4.5 & 0.64 & 0.53      & 0.27    & 0.23 \\ 
512   & 34.0    & 14.7    & 5.3 & 4.5 & 0.64 & 0.53      & 0.21    & 0.23 \\ 
1024  & 48.2    & 20.8    & 5.3 & 4.5 & 0.63 & 0.52      & 0.16    & 0.23 \\ 
2048  & 66.4    & 28.9    & 5.3 & 4.4 & 0.62 & 0.51      & 0.13    & 0.23 \\ 
\hline
\end{tabular}
\vspace*{0.5cm}
\caption[]{Various conformational properties defined in the main text
for 2D polymers at monomer number density $\rho = 7/8$
as a function of chain length $N$.
The aspherity of the chains is characterized by the aspect ratio $\la \lambda_1 \ra/\la \lambda_2 \ra$
and the moments $\Delta_1$ and $\Delta_2$ of the eigenvalues $\lambda_1$ and $\lambda_2$ 
of the inertia tensor. Note that $\Rgyr^2 = \la \lambda_1 \ra + \la \lambda_2 \ra$.
The mean perimeter length $L(N)$ and its relative fluctuation
$\delta L(N)/L(N)$ given in the two last columns
have been obtained assuming a reference distance $a=1.2$
(as indicated in Fig.~\ref{fig_gr}) 
within which at least one monomer from another chain is to be found to 
identify a monomer as a perimeter monomer. 
\label{tab_N}}
\end{table}

\newpage
\clearpage

\newpage
\clearpage

\newpage
\clearpage

\newpage
\clearpage

\newpage
\clearpage

\newpage
\clearpage

\newpage
\clearpage

\newpage
\clearpage

\newpage
\clearpage

\newpage
\clearpage

\newpage
\clearpage

\newpage
\clearpage

\newpage
\clearpage

\newpage
\clearpage

\end{document}